\documentclass[11pt]{article}
\begin{document}

\title{\huge{Local Scale Invariance and General Relativity}} 
\author{Robert Davis Bock}       
\date{\today}
\maketitle
\centerline{\textit{Physics Department, The University of Texas at Austin, Austin, Texas, 78712}}
\centerline{\textit{Present Address:  Georgia Tech Research Institute, Georgia Institute of Technology, Atlanta, Georgia 30332-0834}} 
\begin{abstract}
\noindent
\large{According to the theory of unimodular relativity developed by Anderson and Finkelstein, the equations of general relativity with a cosmological constant are composed of two independent equations, one which determines the null-cone structure of space-time, another which determines the measure structure of space-time.  The field equations that follow from the restricted variational principle of this version of general relativity only determine the null-cone structure and are globally scale-invariant and scale-free.  We show that the electromagnetic field may be viewed as a compensating gauge field that guarantees local scale invariance of these field equations.  In this way, Weyl's geometry is revived.  However, the two principle objections to Weyl's theory do not apply to the present formulation:  the Lagrangian remains first order in the curvature scalar and the non-integrability of length only applies to the null-cone structure.}
\end{abstract}
 
\pagebreak
\section{Introduction}
The theory of general relativity unifies gravitation with the geometry of space-time by replacing the scalar Newtonian gravitational potential with the symmetric metric tensor $g_{\mu\nu}$ of a four-dimensional general Riemannian manifold by means of the equivalence principle.  As is well known, the electromagnetic field may not be interpreted in terms of the geometrical properties of space-time as well.  This difficulty motivated Einstein \cite{EinsteinBOOK,Pais} and many others \cite{Kaluza,Weyl,Klein,Eddington,Schrodinger,Pauli}, immediately following the advent of general relativity, to generalize Riemannian geometry in order to provide a description of electromagnetism within the geometrical framework of space-time.  However, despite tremendous effort, the early unification program was not successful.  

Although modern gauge theory has revealed the common gauge structure of the four fundamental interactions, this problem remains unsolved today.  The symmetry group associated with the gauge theory of gravity\footnote{The local gauge theory of the Poincar\'{e} group is the $U_{4}$ theory of gravity which admits spin and torsion into relativistic gravitational theory \cite{Hehl_1976}.} is the Poincar\'{e} group, the fundamental symmetry group of space-time, while the symmetry group associated with the electromagnetic field is the $U(1)$ group of phase transformations of the wave function, which is an internal or non-geometric \cite{Wigner} symmetry and does not enjoy a space-time interpretation.  Thus, modern gauge theory does not succeed in casting the electromagnetic potentials into the space-time manifold even though it constitutes a great step forward towards the unification of the fields.

In this investigation we show that the electromagnetic field can be introduced as a compensating gauge field that guarantees local scale invariance in general relativity.  There have been a number of scale-invariant theories of gravity proposed in the past.  The first scale-invariant theory of gravity, due to Weyl \cite{Weyl_1918a,Weyl_1918b,Weyl_1919,Weyl_1920, Weyl}, was also an attempt to incorporate electromagnetism into general relativity.  Weyl's theory is based on an elegant generalization of Riemannian geometry that is covariant with respect to both coordinate transformations and local scale transformations.  Since the action that produces Einstein's field equations is only invariant with respect to the former group, Weyl proposed a new action that is invariant with respect to the latter group as well.  This, however, requires a Lagrangian quadratic in the curvature scalar, and therefore leads to field equations that are fourth-order differential equations.  Consequently, Weyl's theory does not reduce to general relativity in the absence of electromagnetism.  Furthermore, Einstein \cite{Einstein_1918} showed that the reading of an atomic clock would depend on its prehistory according to Weyl's theory, which is in conflict with the well-defined electromagnetic spectrum observed from chemical elements.  As a result, Weyl's theory was ultimately rejected.  Years later, Dirac \cite{Dirac_1973} (see also Canuto et al. \cite{Canuto}) revived Weyl's geometry in an attempt to reconcile general relativity with his Large Numbers hypothesis \cite{Dirac_1938}.  Dirac maintains second-order differential equations at the expense of introducing a new scalar field and avoids Einstein's objection with his postulate of a second metric, independent of the gravitational potentials, that determines the interval $ds$ measured by an atomic apparatus.  This theory belongs to a wider class of theories, named variable-gravity theories, that predict a time-dependent variation in the strength of the gravitational interaction.  The advantages and drawbacks of such theories are reviewed by Wesson \cite{WessonBOOK}.  Other attempts at incorporating scale invariance in general relativity (see, for example, Hehl et al. \cite{Hehl_1989}) have been motivated by developments in particle physics.  Since approximate scale invariance has been observed in deep inelastic electron-nucleon scattering \cite{Bjorken_1967,Bjorken_1969} many believe, in accordance with grand unification, that gravitation must also exhibit approximate scale invariance at very high energies.
  
In the following we develop a new method of incorporating local scale invariance into general relativity.  First, we show that a well-known procedure developed by Anderson and Finkelstein \cite{AndersonandFinkelstein} for introducing the cosmological constant removes the scale dependence from the field equations, leaving a set of scale-free field equations behind.  Thus, general relativity with a cosmological constant may be viewed as a union of two independent equations.  One equation determines the null-cone or causal structure of space-time; the other equation determines the measure structure of space-time.  Since the field equations that determine the null-cone structure are globally scale-invariant and scale-free, and are furthermore {\it independent} of the measure equation, we consider them the dynamical equations of a globally scale-invariant theory.  We demand local scale invariance of this theory and see that the electromagnetic field may indeed be treated as a compensating gauge field associated with the group of local scale transformations.  The measure structure is left undetermined by the field equations and is introduced as an external field which is treated as an absolute object.  The theory presented below shares similarities with Weyl's unified theory but does not yield to the same criticisms, since the Lagrangian is first order in the curvature scalar and Einstein's objection does not apply.

\section{Coordinate Invariance vs. Scale Invariance}
Consider an arbitrary action:
\begin{equation}
\label{arbitrary_action}
I=\int W\sqrt{-g}\,d^4x,
\end{equation}
where $W$ is an arbitrary function of the metric tensor and its derivatives.  The variational derivative of $I$ with respect to the metric is defined as:
\begin{equation}
\label{variational_derivative_W}
\frac{\delta I}{\delta g_{\mu\nu}}={\cal W}^{\mu\nu},
\end{equation}
where ${\cal W}^{\mu\nu}$ is a symmetrical contravariant density of the second rank.  As is well known, if the action is invariant under an arbitrary infinitesimal coordinate transformation that vanishes on the boundary:
\begin{equation}
\label{coordinate_transformation}
x^{\prime\mu}=x^{\mu}-\xi^{\mu},
\end{equation}
where $\xi^{\mu}$ are arbitrary infinitesimal functions of the space-time coordinates, then the covariant divergence of ${\cal W}^{\mu\nu}$ vanishes identically:
\begin{equation}
\label{cov_divergence_W}
{\cal W}^{\mu\nu}_{\;\;\; ;\nu}=0.
\end{equation}
This follows from equations (\ref{arbitrary_action}) and (\ref{variational_derivative_W}), noting that the transformation (\ref{coordinate_transformation}) produces a variation in the metric:
\begin{equation}
\label{coordinate_variation_metric}
\delta g_{\mu\nu}=\xi _{\mu ;\nu}+\xi_{\nu ;\mu}.
\end{equation}
Similarly, if the action is invariant under an infinitesimal scale transformation of the metric tensor that vanishes on the boundary:
\begin{equation}
\label{local_scale_transformation}
g_{\mu\nu}\rightarrow \lambda g_{\mu\nu}=(1+\epsilon)g_{\mu\nu},
\end{equation}
where $\lambda =\lambda(x^{\alpha})$ is an arbitrary function of the space-time variables and $\epsilon \ll 1$, then the trace of ${\cal W}^{\mu\nu}$ vanishes identically:
\begin{equation}
\label{trace_W}
{\cal W}^{\mu}_{\;\;\mu}=0.
\end{equation}
This also follows from equations (\ref{arbitrary_action}) and (\ref{variational_derivative_W}), noting that the transformation (\ref{local_scale_transformation}) produces a variation in the metric:
\begin{equation}
\delta g_{\mu\nu}=\epsilon g_{\mu\nu}.
\end{equation}

The action for the gravitational field in the absence of matter is obtained by setting $W=g^{\mu\nu}R_{\mu\nu}$ in (\ref{arbitrary_action}):
\begin{equation}
\label{gravitational_action}
I_{G}=\int g^{\mu\nu}R_{\mu\nu}\sqrt{-g}\,d^4x,
\end{equation}
where $R_{\mu\nu}$ is the Ricci tensor.  The variational derivative of equation (\ref{gravitational_action}) with respect to the metric is:
\begin{equation}
\label{gravitational_variational_derivative}
\frac{\delta I_{G}}{\delta g_{\mu\nu}}=G^{\mu\nu}\sqrt{-g}\equiv \left(R^{\mu\nu}-\frac{1}{2}g^{\mu\nu}R\right)\sqrt{-g},
\end{equation}
Since $R\sqrt{-g}$ is a scalar density, the action (\ref{gravitational_action}) is invariant under the transformation (\ref{coordinate_transformation}).  Therefore, the covariant divergence of $G^{\mu\nu}$ vanishes:
\begin{equation}
\label{covariant_divergence_G}
G^{\mu\nu}_{\;\;\; ;\nu}=0.
\end{equation}
Note that this equation is a consequence of the invariance of the action and is therefore valid for any reasonable metric field distribution $g_{\mu\nu}$, regardless of whether or not $g_{\mu\nu}$ satisifes the field equations.  Equation (\ref{covariant_divergence_G}) also follows from the Bianchi identities.  

While the action (\ref{gravitational_action}) is invariant under general coordinate transformations it is not invariant under the scale transformation (\ref{local_scale_transformation}).  $I_{G}$ is not even invariant under a global scale transformation for which $\lambda=\mbox{constant}$; $R$ and $\sqrt{-g}$ transform under a global scale transformation with Weyl weights $-1$ and $+2$, respectively.  However, the scalar curvature is the only quantity constructed from the metric tensor and its first and second derivatives alone, linear in the latter, that is an invariant under general coordinate transformations.  Therefore, we see that general coordinate invariance and scale invariance of an action of this type are fundamentally incompatible in general relativity.  This is further supported by the fact that the trace of the divergenceless quantity $G^{\mu\nu}$ does not vanish.  Of course, one may proceed as Weyl \cite{Weyl} and consider Lagrangians quadratic in the curvature scalar in order to guarantee scale invariance of the action.  However, the resulting field equations necessarily contain derivatives of the metric tensor higher than the second.  Alternatively, one may proceed as Dirac \cite{Dirac_1973} and introduce a new scalar field that transforms under a scale transformation with Weyl weight $-1$.  This theory has enjoyed only limited success \cite{WessonBOOK}.    
  
\section{Global Scale Invariance and the Cosmological Constant}
Rather than formulating an action principle that is invariant with respect to both coordinate transformations and scale transformations simultaneously, we reformulate general relativity so that the scale-dependent quantity, the Ricci scalar curvature, remains undetermined by the field equations themselves.  As a result, the remaining field equations become scale-free.  This allows us to treat these equations as the dynamical equations of a globally scale invariant theory that can be gauged locally. 
  
First, let us consider Einstein's equations in the absence of matter:
\begin{equation}
\label{Einstein_field_equations_free}
R_{\mu\nu}-\frac{1}{2}Rg_{\mu\nu}=0,
\end{equation}
which follow from the variational principle $\delta I_G=0$, in which the metric components are varied independently.  While the gravitational action is not invariant under a global scale transformation defined by equation (\ref{local_scale_transformation}) with $\lambda=\mbox{constant}$, Einstein's free-field equations are invariant with respect to global scale transformations.  This follows because $R_{\mu\nu}$, $R$, and $g_{\mu\nu}$ transform under a global scale transformation with Weyl weights $0$, $-1$, and $+1$ respectively.  The fact that the equations are globally scale invariant does not imply that the theory is also scale-free.  This follows by taking the trace of (\ref{Einstein_field_equations_free}), giving:
\begin{equation}
\label{Ricci_scalar_vanishes}
R=0.
\end{equation}
Because $R$ vanishes, pure gravity is also scale-free:  pure gravity contains no intrinsic length scale.  Note that equation (\ref{Ricci_scalar_vanishes}) is not independent of (\ref{Einstein_field_equations_free}); rather, it is a consequence of the field equations. 

We stress that the terms scale-free and scale-invariant are similar but not identical.  A theory is scale-free if it does not contain any constant fundamental length scale.  A theory is (globally) locally scale-invariant if, in addition to the absence of any fundamental length scale, the dynamical equations are covariant with respect to (global) local scale transformations.  Note that a theory may be scale-free and not scale-invariant.  As we saw above, pure gravity is globally scale-invariant:  the equations of pure gravity are covariant, in fact invariant, with respect to global scale transformations, and since $R$ vanishes pure gravity is also scale-free.
      
Once matter is introduced, global scale invariance of the theory is lost.  The action for the gravitational field in the presence of matter is:
\begin{equation}
I=I_{G}+I_{M},
\end{equation}
where 
\begin{equation}
\label{matter_action}
I_{M}=-2\kappa\int L_{M}\sqrt{-g}\,d^4x,
\end{equation}
is the matter action and $\kappa=\frac{8\pi G}{c^4}$ is the Einstein gravitational constant.  The resulting field equations are: 
\begin{equation}
\label{Einstein_field_equations_matter}
R_{\mu\nu}-\frac{1}{2}Rg_{\mu\nu}=\kappa T_{\mu\nu}.
\end{equation}
Taking the trace of the above equation yields:  $R=-\kappa T$.  Again, this equation is contained in the field equations.  The equations (\ref{Einstein_field_equations_matter}) may be considered globally scale invariant if one assumes that the product $\kappa T_{\mu\nu}$ is scale-invariant \cite{Canuto}, regardless of the manner in which each term transforms individually.  However, since $R$ does not vanish the theory is no longer globally scale invariant.  Rest masses introduce an intrinsic length scale. 

There is a way of reformulating the theory so that the scale dependence remains undetermined by the field equations themselves.  This is accomplished by a well-known procedure developed by Anderson and Finkelstein \cite{AndersonandFinkelstein} for introducing the cosmological constant into Einstein's equations, not as a predetermined coefficient of the action, but as an arbitrary integration constant.  Indeed, if one introduces the constraint in the variational principle:
\begin{equation}
\label{constraint}
\sqrt{-g}=\sigma(x),
\end{equation}
where $\sigma(x)$ is a scalar density of weight $+1$, an external field provided by nature, then the components of the metric tensor cannot be varied independently in the action principle, but must satisfy:
\begin{equation}
\label{determinant_constant}
\delta \sqrt{-g}=-\frac{1}{2}\sqrt{-g}g_{\mu\nu}\delta g^{\mu\nu}=0.
\end{equation}
The resulting field equations express the equality of the traceless parts of equation (\ref{Einstein_field_equations_matter}):
\begin{equation}
\label{traceless_field_equations_matter}
R_{\mu\nu}-\frac{1}{4}g_{\mu\nu}R=\kappa\left(T_{\mu\nu}-\frac{1}{4}g_{\mu\nu}T\right).
\end{equation}
Because of equation (\ref{covariant_divergence_G}) and 
\begin{equation}
T^{\mu\nu}_{\;\;\; ;\nu}=0,
\end{equation}
one obtains:
\begin{equation}
R+\kappa T=4\Lambda,
\end{equation}
where $\Lambda$ is an integration constant and the factor four is introduced for convenience.  Substituting this back into the field equations (\ref{traceless_field_equations_matter}) we recover Einstein's field equations with a cosmological constant:
\begin{equation}
\label{Einstein_field_equations_cosmological}
R_{\mu\nu}-\frac{1}{2}Rg_{\mu\nu}+\Lambda g_{\mu\nu}=\kappa T_{\mu\nu}.
\end{equation}
Einstein \cite{Einstein_1919} examined the field equations (\ref{traceless_field_equations_matter}) with $T_{\mu\nu}$ representing only the stress-energy tensor of the electromagnetic field, and similarly recovered the cosmological constant as a constant of integration.  Anderson and Finkelstein \cite{AndersonandFinkelstein} were the first to propose the above general procedure in their theory of unimodular relativity.  This formulation has the attractive property that the contribution of vacuum fluctuations automatically cancels on the right hand side of equation (\ref{traceless_field_equations_matter}) \cite{Weinberg}.  The full theory is contained in either equations (\ref{constraint}) and (\ref{traceless_field_equations_matter}) or equations (\ref{constraint}) and (\ref{Einstein_field_equations_cosmological}).  The full theory is not scale-invariant, because it contains the constraint (\ref{constraint}), which manifests itself in the field equations by the presence of the fundamental length $\Lambda^{-1/2}$.  However, the set of equations (\ref{traceless_field_equations_matter}) are scale-free.  Equation (\ref{Einstein_field_equations_cosmological}) is valid for any value of $\Lambda$ which is an arbitrary constant of integration.  The condition (\ref{constraint}) does not determine the value of $\Lambda$, which must be determined by external conditions.

The ability to remove the scale dependence from the field equations is a consequence of the ability to reduce the metric tensor into two nontrivial geometric objects \cite{AndersonandFinkelstein}:  $g$ the determinant of $g_{\mu\nu}$, and $\gamma_{\mu\nu}$ the relative tensor $g_{\mu\nu}/(\sqrt{-g})^{1/2}$ of determinant $-1$.  The determinant determines entirely the measure structure of space-time, while the relative tensor alone determines the null-cone or causal structure.  In unimodular relativity, the irreducible relative tensor $\gamma_{\mu\nu}$ is the fundamental geometric object of space-time.  The metric tensor: 
\begin{equation}
\label{g_identification}
g_{\mu\nu}=(\sqrt{-g})^{1/2}\gamma_{\mu\nu},
\end{equation}
is treated as an artificial construct of two independent entities, the fundamental object $\gamma_{\mu\nu}$ and the measure field $\sqrt{-g}$.  The measure field is only included in the formulation of the action principle in order to maintain general covariance.  Because of the constraint (\ref{constraint}), the invariance group of unimodular relativity is the subgroup of the Einstein group with unit determinant:
\begin{equation}
\mbox{det}\left|\frac{\partial x^{\prime \mu}}{\partial x^{\nu}}\right|=1.
\end{equation}
(See Reference \cite{AndersonBOOK} for a lucid discussion of the terms ``invariance'' and ``covariance'' as they are used here.) 
     
\section{Geometry and Space-Time Measurements}
The bifurcation of general relativity into two independent parts suggests a new way of looking at the connection between geometry and space-time measurements.  In general relativity, actual space-time is represented geometrically by a Riemannian manifold ${\cal R}$:  there exists a transparent correspondence between geometrical quantities on the one hand and physical space-time measurements on the other hand.  The square of the length of an arbitrary vector $A^{\mu}$ in ${\cal R}$ is:
\begin{equation}
\label{A_length_GR}
A^2=g_{\alpha\beta}A^{\alpha}A^{\beta},
\end{equation}
and the change of an arbitrary vector $A^{\mu}$ under an infinitesimal displacement $dx^{\alpha}$ in ${\cal R}$ is:
\begin{equation}
\label{parallel_displacement}
dA^{\mu}=-\Gamma^{\mu}_{\rho\sigma}A^{\rho}dx^{\sigma},
\end{equation}
where $\Gamma^{\mu}_{\rho\sigma}$ is the Christoffel symbol of the second kind:
\begin{equation}
\label{Christoffel_symbol}
\Gamma^{\mu}_{\rho\sigma}=\frac{g^{\mu\alpha}}{2}\left(\frac{\partial g_{\alpha\rho}}{\partial x^{\sigma}}+\frac{\partial g_{\alpha\sigma}}{\partial x^{\rho}}-\frac{\partial g_{\rho\sigma}}{\partial x^{\alpha}}\right).
\end{equation}
Equation (\ref{Christoffel_symbol}) is obtained from the condition:
\begin{equation}
\label{covariant_derivative_metric}
g_{\mu\nu ;\lambda}=0,
\end{equation}
which follows from the requirement that the length of an arbitrary vector is preserved under parallel displacement in $\cal{R}$.  Because we identify ${\cal R}$ with physical space-time in general relativity it follows that the quantity $A^{2}$ may be identified with the result of a physical space-time measurement.  Furthermore, it follows that the parallel displacement of a vector $A^{\mu}$ in ${\cal R}$ may be equated with the transfer of the corresponding physical length in actual space-time.  Moreover, any generalization of the geometrical manifold ${\cal R}$ will presumably manifest itself as a generalization of the behavior of physical rods and clocks.  These assumptions are fundamental to Einstein's theory.

Unimodular relativity, owing to the bifurcation of the metric tensor, admits a substructure to the manifold ${\cal R}$, and hence permits the introduction of another geometrical manifold that is not directly related to space-time measurements.  To see this we rewrite equation (\ref{A_length_GR}) in terms of the quantities $\gamma_{\alpha\beta}$ and $\sqrt{-g}$ of unimodular relativity:
\begin{equation}
\label{A_length_UR}
A^2=(\sqrt{-g})^{1/2}\gamma_{\alpha\beta}A^{\alpha}A^{\beta}.
\end{equation}
We define the ``length'':
\begin{equation}
\label{a}
a^2\equiv\gamma_{\alpha\beta}A^{\alpha}A^{\beta},
\end{equation}
so that equation (\ref{A_length_UR}) becomes:
\begin{equation}
\label{A_length_a}
A^2=(\sqrt{-g})^{1/2}a^{2}=\sigma^{1/2}a^{2}.
\end{equation}
We see that a physical measurement is obtained by multiplying two independent quantities, $\sigma^{1/2}$ and $a^2$.  In general relativity, both of these quantities are obtained from the geometrical manifold ${\cal R}$.  However, since $\sigma$ and $\gamma_{\mu\nu}$ are completely independent in unimodular relativity, we may construct a sub-geometry that is only associated with the fundamental object $\gamma_{\mu\nu}$.  Thus, we define a manifold ${\cal M}$.  On this manifold we define a metric tensor $\tilde{g}_{\mu\nu}$ and an affine connection $\tilde{\Gamma}^{\mu}_{\rho\sigma}$.  We do not assume that $\tilde{g}_{\mu\nu}$ satisfies equation (\ref{covariant_derivative_metric}); the quantities $\tilde{\Gamma}^{\mu}_{\rho\sigma}$ are not necessarily Christoffel symbols.  Furthermore, the measure field $\sqrt{-\tilde{g}}$ is not necessarily identified with $\sigma$.  A correspondence with physical space-time measurements may be obtained from ${\cal M}$ if we make the following identification: 
\begin{equation}
\label{gamma_identification}
\gamma_{\mu\nu}\equiv\frac{\tilde{g}_{\mu\nu}}{(\sqrt{-\tilde{g}})^{1/2}}.
\end{equation}
Thus, the Riemannian manifold ${\cal R}$ in unimodular relativity may be treated as an artificial construct defined from the manifold ${\cal M}$ by the relationship:
\begin{equation}
\label{Riemannian_manifold}
g_{\mu\nu}=\sigma^{1/2}\gamma_{\mu\nu}=\sigma^{1/2}\frac{\tilde{g}_{\mu\nu}}{(\sqrt{-\tilde{g}})^{1/2}}.
\end{equation}  
  
In the original formulation of unimodular relativity \cite{AndersonandFinkelstein}, Anderson and Finkelstein tacitly assumed that the manifold ${\cal M}$ was also a Riemannian manifold.  However, a physical measurement defined in this manner admits a natural generalization, for the only geometrical quantity obtained from ${\cal M}$ is the scale-independent quantity $a^{2}$.  Consequently, the choice of the scale of the metric tensor on ${\cal M}$ is arbitrary.  Therefore, we may choose a Weyl manifold ${\cal W}$ for ${\cal M}$.  The law of parallel displacement of an arbitrary vector $A^{\mu}$ in ${\cal W}$ is:
\begin{equation}
\label{W_parallel_displacement}
dA^{\mu}=-\tilde{\Gamma}^{\mu}_{\rho\sigma}A^{\rho}dx^{\sigma},
\end{equation}
where $\tilde{\Gamma}^{\alpha}_{\beta\gamma}$ is the Weyl affine connection:
\begin{equation}
\label{Weyl_affine_connection}
\tilde{\Gamma}^{\alpha}_{\beta\gamma}=\Gamma^{\alpha}_{\beta\gamma}-\tilde{g}^{\sigma\alpha}[\tilde{g}_{\sigma\beta}\varphi_{\gamma}+\tilde{g}_{\sigma\gamma}\varphi_{\beta}-\tilde{g}_{\beta\gamma}\varphi_{\sigma}].
\end{equation}
$\Gamma^{\alpha}_{\beta\gamma}$ now represents the Christoffel symbol constructed from the quantities $\tilde{g}_{\mu\nu}$.  The Weyl affine connection $\tilde{\Gamma}^{\alpha}_{\beta\gamma}$ follows from $\Gamma^{\alpha}_{\beta\gamma}$ by the substitution:
\begin{equation}
\partial_{\gamma}\rightarrow\partial_{\gamma}-2\varphi_{\gamma}.
\end{equation}
The vector $\varphi_{\mu}$ serves as the connection coefficient for the parallel displacement of length in ${\cal W}$:
\begin{equation}
\label{W_l_parallel_displacement}
dl=+\varphi_{\beta}dx^{\beta}l,
\end{equation}
where $l$ is the length of an arbitrary vector in ${\cal W}$.  As long as physically measured quantities are associated with the manifold ${\cal W}$ via equation (\ref{Riemannian_manifold}), the comparison of physical lengths at different points in space-time is an unambiguous procedure that is not to be confused with the comparison of vector lengths at different points in the manifold ${\cal W}$.  Consequently, the identification ${\cal M}={\cal W}$ is not incompatible with the existence of the well-defined electromagnetic spectrum observed from chemical elements and Einstein's objection does not apply to {\it this} use of a Weyl geometry.  Both Weyl \cite{Weyl} and Eddington \cite{Eddington} envisioned that such a geometry could be constructed which is not immediately identifiable with actual space-time but could be associated with physical measurements.  We see that unimodular relativity provides a natural framework for the realization of this vision.  
 
\section{Local Scale Invariance and the Electromagnetic Field}
The field equations of unimodular relativity (\ref{traceless_field_equations_matter}) are the equations for the quantities $\gamma_{\mu\nu}$ corresponding to the special case ${\cal M}={\cal R}$.  These equations are globally scale-invariant and scale-free, and are furthermore independent of the measure equation.  Therefore, we view these equations as the dynamical set of equations of a globally scale-invariant theory.  This interpretation is further supported by the fact that equation (\ref{traceless_field_equations_matter}) is traceless (see equation (\ref{trace_W})). 

We now demand local scale invariance of this globally scale-invariant theory.  We replace the Ricci tensor $R_{\mu\nu}$ in the action (\ref{gravitational_action}) by the scale-invariant Ricci tensor $\tilde{R}_{\mu\nu}$ of Weyl's theory:
\begin{equation}
\label{Weyl_ricci_tensor}
\tilde{R}_{\alpha\beta}=\frac{\partial \tilde{\Gamma}^{\rho}_{\alpha\beta}}{\partial x^{\rho}}-\frac{\partial \tilde{\Gamma}^{\rho}_{\alpha\rho}}{\partial x^{\beta}}+ \tilde{\Gamma}^{\sigma}_{\alpha\beta} \tilde{\Gamma}^{\rho}_{\rho\sigma}-\tilde{\Gamma}^{\sigma}_{\alpha\rho}\tilde{\Gamma}^{\rho}_{\beta\sigma},
\end{equation}
where the quantities $\tilde{\Gamma}^{\alpha}_{\beta\gamma}$ are constructed from the metric tensor and the vector field $\varphi_{\alpha}$ according to equation (\ref{Weyl_affine_connection}).  Under the transformation (\ref{local_scale_transformation}), $\varphi_{\alpha}$ transforms according to:
\begin{equation}
\label{vector_gauge_transformation}
\varphi_{\alpha}\rightarrow\varphi_{\alpha}+\frac{1}{2}(\mbox{log}\, \lambda)_{,\alpha}=\varphi_{\alpha}+\frac{1}{2}\frac{\epsilon_{,\alpha}}{\epsilon},
\end{equation}
where a comma denotes ordinary differentiation.  Equation (\ref{vector_gauge_transformation}) guarantees the invariance of $\tilde{\Gamma}^{\alpha}_{\beta\gamma}$ under local scale transformations.  Thus, $\varphi_{\alpha}$ may be considered a compensating gauge field that guarantees local scale invariance.  Similarly, $\tilde{R}_{\alpha\beta}$ is an invariant under transformation (\ref{local_scale_transformation}) and the scalar curvature $\tilde{R}$:
\begin{equation}
\label{Weyl_scalar_curvature}
\tilde{R}=R+6(\varphi^{\alpha}\varphi_{\alpha})-6\varphi^{\alpha}_{\,\,;\alpha},
\end{equation}
transforms with Weyl weight $-1$.  The free Lagrangian $L_{0}$ for the gauge field $\varphi_{\mu}$ is the lowest order covariant combination of the gauge potentials:
\begin{equation}
L_{0}=-\frac{1}{16\pi}f_{\mu\nu}f^{\mu\nu},
\end{equation}
where $f_{\mu\nu}=\varphi_{\mu,\nu}-\varphi_{\nu,\mu}$ and the indices are raised with the metric $\tilde{g}^{\mu\nu}$.  Consequently, the action is:
\begin{equation}
\label{scale_invariant_gravitational_action}
\int\left[R+6(\varphi^{\alpha}\varphi_{\alpha})-6\varphi^{\alpha}_{\,\,;\alpha}-\frac{k}{16\pi}f_{\mu\nu}f^{\mu\nu}\right]\sqrt{-\tilde{g}}\,d^4x+I_{M},
\end{equation}
where $k$ is a constant that transforms under a local scale transformation with Weyl weight $+1$.  The field equations from the restricted variation of unimodular relativity are:
\begin{equation}
\label{unified_field_equations}
\tilde{R}_{\mu\nu}-\frac{1}{4}\tilde{R}\tilde{g}_{\mu\nu}=k T^{(EM)}_{\mu\nu}+\kappa\left(T_{\mu\nu}-\frac{1}{4}\tilde{g}_{\mu\nu}T\right),
\end{equation}
where $T^{(EM)}_{\mu\nu}\equiv \frac{1}{8\pi}\left(f_{\mu\alpha}f_{\nu}^{\;\;\alpha}-\frac{1}{4}\tilde{g}_{\mu\nu}f^{\alpha\beta}f_{\alpha\beta}\right)$.  Note that the covariant derivative of the metric tensor does not vanish and therefore the integral over $\delta R_{\mu\nu}$ does not reduce to a surface integral.  These field equations are invariant under local scale transformations (\ref{local_scale_transformation}).  They are similar, but not identical to the Maxwell-Einstein system of equations.  The correction terms are on the order of the cosmological constant.  Bergmann and Einstein \cite{Bergmann} have examined the set of equations:  $\tilde{R}_{\mu\nu}-\frac{1}{2}\tilde{g}_{\mu\nu}\tilde{R}=0$, and found that its solutions do not satisfy reasonable boundary conditions.  However, they identified $\tilde{g}_{\mu\nu}$ with the scale-dependent metric tensor of space-time and also failed to include the term $k T_{\mu\nu}^{(EM)}$.

The field equations for the quantities $\varphi_{\mu}$ are obtained by a variation of $\varphi_{\mu}$ in (\ref{scale_invariant_gravitational_action}).  This produces Maxwell's free-field equations:
\begin{equation}
f^{\mu\nu}_{\;\;\; ;\nu}=0,
\end{equation}
only for the case of a vanishing cosmological constant.  Therefore, we may identify $\varphi_{\mu}$ as being proportional to the electromagnetic four-vector potential, with the condition that exact gauge invariance of electromagnetism is connected to a vanishing cosmological constant.

\section{Discussion}
We have used the mathematical formalism developed by Weyl, originally used unsuccessfully to generalize Einstein's relativity, to generalize unimodular relativity.  We may adopt Weyl's formalism in unimodular relativity because the fundamental geometrical object is not the metric tensor $g_{\mu\nu}$, but the scale-invariant relative tensor $\gamma_{\mu\nu}$.  Consequently, the electromagnetic field may be viewed as a compensating gauge field that guarantees local scale invariance of the field equations in unimodular relativity. 

In the limit that the measure field may be ignored, viz. small length scales, the full theory is scale invariant, in accordance with the belief that general relativity should exhibit approximate scale-invariance at high energies.  The source of the measure field $\sigma(x)$ was not specified in the original paper \cite{AndersonandFinkelstein}; its value was simply provided by an external condition.  We postulate that the source of the measure field is the background mass distribution of the distant stars, and consequently the tensor $T_{\mu\nu}$ only represents local matter.  Note that in Einstein's original formulation of general relativity there is no distinction between a local mass distribution, such as a planet or the sun, and the background mass distribution of the distant stars.  The entire matter content of the universe is contained in the matrix $T_{\mu\nu}$.  However, if we postulate that the distant stars are the source of the field $\sigma(x)$ then such a distinction can be made.  According to this postulate, distant matter would determine the measure structure of space-time and local matter would detemine the null-cone structure of space-time.  Since the volume of space-time within the interval $d^{4}x$ is $\sigma (x)d^{4}x$, the very existence of the volume element would then be tied into the boundary conditions defined by the distant stars.  

This postulate solves an important problem concerning Mach's principle in general relativity.  Mach's principle states that inertia cannot be defined relative to absolute space, but must be defined relative to the entire matter content of the universe.  As is well known, one of the reasons Einstein \cite{Einstein_1917} introduced the cosmological constant into the gravitational field equations was to accommodate Mach's principle.  Einstein hoped that his reformulation of the field equations with a cosmological constant would eliminate solutions in the absence of mass, giving $g_{\mu\nu}=0$ when $T_{\mu\nu}=0$.  Soon afterward, de Sitter \cite{deSitter} showed that this was not the case; a solution with a non-zero $g_{\mu\nu}$ existed even in the absence of matter.  However, if the above postulate is adopted then de Sitter's solution is not in conflict with Mach's principle, for then the condition $T_{\mu\nu}=0$ would only indicate the absence of a local mass distribution.  The existence of de Sitter's solution would then be connected to a non-zero $\sigma(x)$, which would presumably vanish if the background mass distribution of the distant stars were to disappear.  Note that this solution to the problem of Mach's principle in general relativity is similar to that provided by Brans and Dicke \cite{Brans_Dicke}, who supplied a scalar field in addition to the metric tensor.  However, instead of adding an additional degree of freedom, we identify one of the ten degrees of freedom of the metric tensor as a scalar field connected to the boundary conditions of space-time.

The theory outlined above may provide insight into the strange behavior of quantum particles.  As is well known, singularities of the field $g_{\mu\nu}$ traverse geodesics of the manifold, which are identified with the trajectories of material particles.  However, according to the theory described above, this is just an approximation, for material particles should be viewed as singularities of the field $\gamma_{\mu\nu}$, not $g_{\mu\nu}$.

\section{Acknowledgments}
I would like to thank Professor W. Schieve and Professor L. Horwitz for their input.  In addition, I would like to thank Dr. L. Bardenshteyn for reading this manuscript.


\end{document}